\UseRawInputEncoding
\documentclass[pdflatex,sn-mathphys-num]{sn-jnl}

\usepackage{graphicx}%
\usepackage{multirow}%
\usepackage{amsmath,amssymb,amsfonts}%
\usepackage{amsthm}%
\usepackage{mathrsfs}%
\usepackage[title]{appendix}%
\usepackage{xcolor}%
\usepackage{textcomp}%
\usepackage{manyfoot}%
\usepackage{booktabs}%
\usepackage{algorithm}%
\usepackage{algorithmicx}%
\usepackage{algpseudocode}%
\usepackage{listings}%

\usepackage{float}
\usepackage{graphicx}
\usepackage{dcolumn}
\usepackage{bm}
\usepackage{soul}
\usepackage{hyperref}
\usepackage[hang,flushmargin]{footmisc}
\usepackage{changes}
\definechangesauthor[name=kdb, color=red]{kdb}
\makeatletter
\def\blfootnote{\xdef\@thefnmark{}\@footnotetext}
\makeatother

\newcommand{\fluenceUnitsJ}{ J/cm$^{2}$}
\newcommand{\titanate}{CaTiO$_3$}

\newcommand{\ms}{m/s }

\begin{document}

\title[Article Title]{Absorption Spectroscopy of \texorpdfstring{$^{40}$Ca}{} Atomic Beams Produced via Pulsed Laser Ablation: A Quantitative Comparison of Ca and \texorpdfstring{\titanate}{} Targets}

\author*[1]{\fnm{Kevin D.} \sur{Battles}}\email{kevin.battles@gtri.gatech.edu}

\author[1]{\fnm{Brian J.} \sur{McMahon}}

\author[1]{\fnm{Brian C.} \sur{Sawyer}}

\affil*[1]{\orgname{Georgia Tech Research Institute},\orgaddress{ \city{Atlanta}, \postcode{30318}, \state{GA}, \country{USA}}}

\abstract{Pulsed laser ablation is an increasingly prevalent method for fast ion trap loading of various species, however characteristics of the ablation target source material can affect the ion-loading process. One factor which can reduce the atomic flux from a target is oxidation during atmospheric exposure when preparing or making changes to the ion trap vacuum system. Recent work has shown that perovskite ablation targets produce consistent atomic densities even after exposure to atmosphere when compared to elemental source targets. In this work, we directly compare calcium (Ca) and calcium-titanate (\titanate) ablation targets, characterizing the neutral atomic beam flux using resonant, time-resolved absorption spectroscopy of the 423~nm $^{1}S_0 \rightarrow$ $^{1}P_1$ transition in neutral Ca. We measure the ablation plume longitudinal and transverse temperatures, number density, ion production, and spot lifetime for each target. In addition, we compare the probe laser beam absorption for both targets before and after 21 hours of exposure to atmosphere, demonstrating the relative robustness of the \titanate~source.}
\makeatletter\def\Hy@Warning#1{}\makeatother
\blfootnote{This work relates to Department of Navy award N00014-20-1-2427 issued by the Office of Naval Research. The United States Government has a royalty-free license throughout the world in all copyrightable material contained herein.}

\maketitle

\section{Introduction}
Trapped ions are promising candidates for quantum information science due to their high fidelity operations \cite{an_high_2022,clark_high-fidelity_2021, srinivas_high-fidelity_2021,ballance_high-fidelity_2016,gaebler_high-fidelity_2016} and long trap lifetimes \cite{wang_evolution_2021,wu_increase_2021}. Trapped-ion loading has typically been done by using a resistively heated oven in proximity to the trapping region, which contains an elemental atom source to produce a thermal beam of neutral precursor atoms~\cite{kjaergaard_isotope_2000,gulde_simple_2001}. This oven loading process can take from tens of seconds to a few minutes, and heating of the oven to 400-600 $^{\circ}$C can cause residual heating of the ion trap and surrounding vacuum system. Also, if vacuum system changes are needed, oxidation of the oven elemental source metal can reduce the lifetime of the oven.  Other innovative methods of ion loading have been investigated such as pulsed laser ablation\cite{leibrandt_laser_2007,vrijsen_efficient_2019, sameed_ion_2020,white_isotope-selective_2022,shi_ablation_2023}, optically heated atomic sources \cite{gao_optically_2021} and micro-fabricated hotplates \cite{schwindt_highly_2016}. However the optically heated sources currently have large optical power requirements ($\sim$ 1-2 W) to reach required densities of atoms for certain species \cite{gao_optically_2021} and micro-fabricated hotplates tend to have reduced trapped ion lifetimes due to baking temperature constraints \cite{schwindt_highly_2016}.

Increasing numbers of researchers are moving towards laser ablation loading of trapped ions for many practical reasons including fast loading times \cite{shi_ablation_2023,vrijsen_efficient_2019} and elimination of residual trap heating. Some commonly-used ablation techniques for ion trap loading are photoionization of neutral atoms within the ion trap \mbox{\cite{shi_ablation_2023,white_isotope-selective_2022,hannig_towards_2019,vrijsen_efficient_2019,shao_laser_2018,sheridan_all-optical_2011,hendricks_all-optical_2007}} or direct ion loading \mbox{\cite{sameed_ion_2020,zimmermann_laser_2012}}, which several groups have demonstrated  for various species in RF and Penning ion traps. A primary concern in laser ablation loading of ions is the atom yield per ablation pulse and the number density stability versus ablation pulse number (i.e. spot lifetime) and thus how often the ablation laser spot must be re-positioned on the target. As with oven sources, oxidation of the target sample will reduce the available atom number density (e.g. during prolonged trap assembly or subsequent vacuum system breaks). Several groups have shown ablation loading targets to have relatively long spot lifetimes of tens of thousands of pulses given laser fluences in the $\sim$ 0.1 - 3.0 \fluenceUnitsJ\ range\cite{shi_ablation_2023,white_isotope-selective_2022,sameed_ion_2020,olmschenk_laser_2017,sheridan_all-optical_2011,hendricks_all-optical_2007}. Additionally, laser ablation targets of different salt and perovskite substrates have been shown to produce a consistent ion yield when compared to elemental sources which are subject to oxidation when exposed to atmosphere \cite{osada_deterministic_2022,osada_feasibility_2022,olmschenk_laser_2017}.

In this work, we characterize $^{40}$Ca neutral atom plumes produced from pulsed laser ablation of both Ca and \titanate\ targets for future integration with, for example, a compact Penning ion trap \cite{mcmahon_doppler-cooled_2020,mcmahon_second-scale_2022,mcmahon_individual-ion_2024}. Calcium metal is highly reactive with oxygen and an oxidative layer is formed when the metal is exposed to atmosphere. However, \titanate\ is chemically inert and does not degrade under air exposure, making it a potential alternative as a $^{40}$Ca ion source in ion-trapping experiments. In this work, we perform a comparative characterization of the neutral atomic flux produced through pulsed laser ablation of Ca and \titanate\ targets using absorption spectroscopy of the 423 nm $^{1}S_0 \rightarrow ^{1}P_1$ transition in neutral Ca. We derive an expression for the neutral atomic density extracted from resonant laser beam absorption measurements, giving quantitative comparisons of the ablation plume temperature and spot lifetimes under vacuum as well as the probe laser beam absorption for both targets before and after exposure to atmosphere. We measure the time-dependent phase space profiles of ablation plumes in the absence of restrictive trap electrode geometries. Combined with photoionization or electron impact ionization cross sections, our temperature and density measurements permit quantitative estimates of $^{40}$Ca ablation trap loading rates for arbitrary trap geometries.
\section{Experiment}\label{sec:Experiment}
\begin{figure}
    \centering
    \includegraphics[width=\columnwidth]{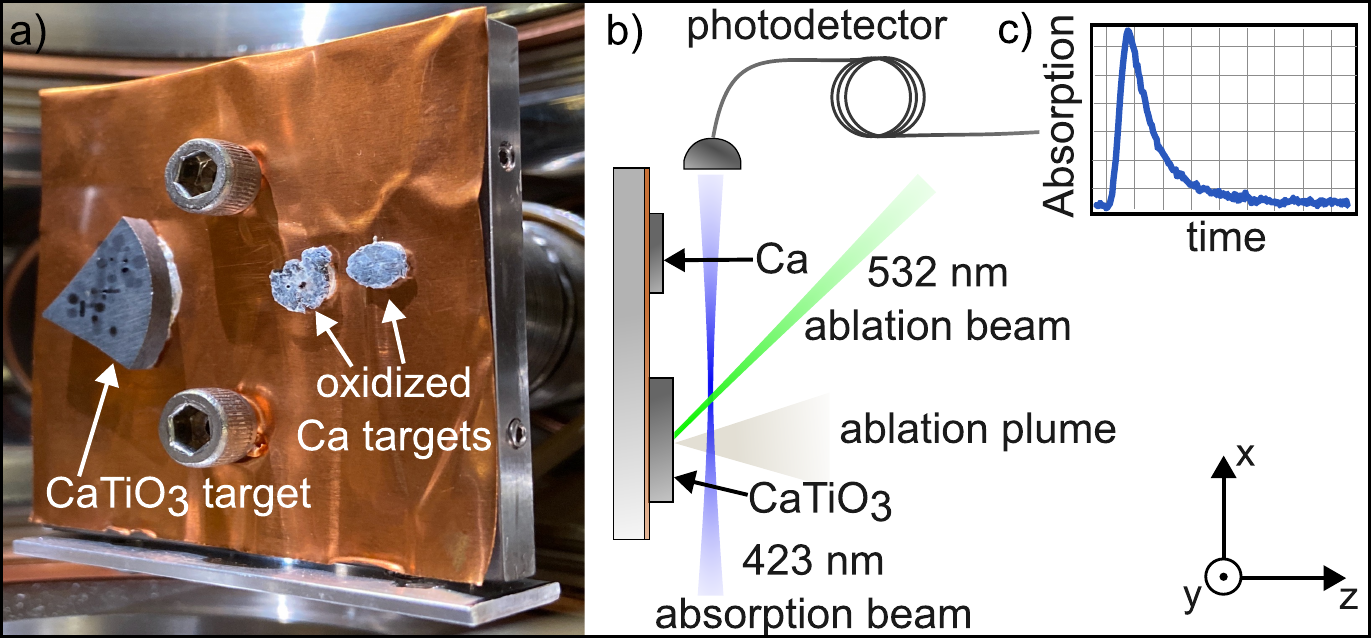}
    \caption{a) \titanate~(left) and Ca (right) ablation targets mounted in the test vacuum chamber. This photo was taken after $\sim$ 21-h of atmospheric exposure, and oxidation is clearly visible on the two Ca targets. b) Top-down-view illustration of the experimental setup and c) a typical time-resolved probe laser absorption signal.}
    \label{fig:ExperimentalLayout}
\end{figure}

The experiment setup is shown in Fig.~\ref{fig:ExperimentalLayout}, and consists of Ca and \titanate\ ablation targets fixed to a mounting plate on a linear motion feedthrough with ultrahigh vacuum (UHV) compatible conductive epoxy. The ablation targets are mounted in a 6-inch stainless steel spherical octagon vacuum chamber pumped to a pressure of $6 \times 10^{-7}$ Torr. The \titanate\ target is a commercial sputtering target disk (1 in. diameter, 0.125 in. thick) that was segmented into a smaller ($\sim 0.3 \textrm{ in}^2$) piece. The Ca targets are commercial 99.5\% pure Ca metal granules pressed into a disk shape to increase the ablation surface area. The targets are mounted with their surfaces at 45$^{\circ}$ to the ablation laser beam (see Fig.~\ref{fig:ExperimentalLayout}b). For ablation, we use a commercial 532 nm Q-switched laser with a pulse width of $\sim5$~ns, pulse repetition rate of 1-15 Hz, and beam waist of 360 $\mu$m at the target locations. The absorption probe laser beam is a CW, 423-nm external cavity diode laser (ECDL) which is tuned to be resonant with the $^{1}S_0 \rightarrow\ ^{1}P_1$ transition in $^{40}$Ca with a 280~$\mu$m waist and 200~$\mu$W of optical power. The probe laser beam fiber launch is mounted on a 2-axis motorized stage and transmitted through the vacuum chamber view port along the $x$-axis parallel to the ablation target surfaces. The probe beam was placed at initial $z$-position of approximately 4~mm from the surface of the Ca and \titanate\ targets. The time-resolved probe beam transmission is measured with a photodetector connected to an oscilloscope, which is triggered simultaneously with the ablation pulse.

\section{Ablation Characterization}\label{sec:AblationCharacterization}
 \begin{figure*}
    \centering
    \includegraphics[width=\textwidth]{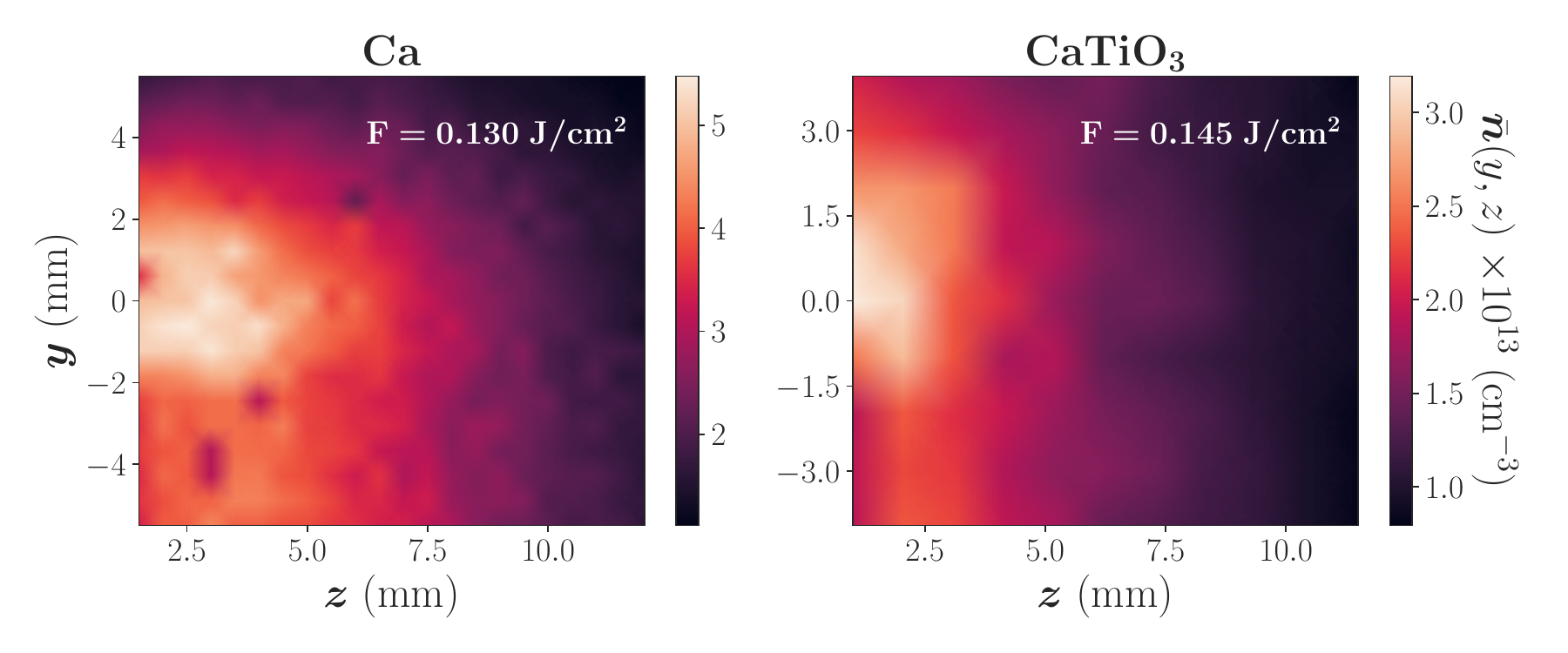}
    \caption{Ablation plume images taken by scanning the 423 nm probe laser beam in the $yz$-plane while pulsing the 532 nm ablation beam. The 423 nm transmission is measured at each $(y,z)$ position, allowing computation of the time-averaged density profile, $\bar{n}(y,z)$, at a fixed time after ablation.}
    \label{PlumeImages}
\end{figure*}
We present an analytic derivation of the spatially-dependent atomic beam number density, $n$, of the $^{1}S_0$ ground state atoms as a function of fractional probe laser beam absorption. Here we neglect leakage from the excited $^{1}P_1$ state to the metastable $^{1}D_2$ state, which has a lifetime of $\sim$ 2~ms \cite{pasternack_experimental_1980,fischer_allowed_2003}, resulting in a large branching ratio of the $\left( ^{1}P_1 \rightarrow\ ^{1}S_0\ \right):\ \left( ^{1}P_1 \rightarrow\ ^{1}D_2\right)$ transitions of $\sim$ $1.0(2)\times10^{5}$
 \cite{beverini_measurement_1989}.
 We define the absorption coefficient, $\alpha$, of the 423~nm probe beam through the ablation plume, assuming the atom density is spatially dependent such that
\begin{align}\label{eq:alpha1}
\alpha (x,y,z) &= k_P\frac{2n(x,y,z)\mu_{eg}}{\epsilon_0 E_P}\int P(v_x)\mathfrak{I}(\rho_{01})dv_x\\
&= k_P\frac{2n(x,y,z)\mu_{eg}^2}{\epsilon_0 \hbar\Omega_p}\int P(v_x)\mathfrak{I}(\rho_{01})dv_x
\end{align}
where $k_P=\frac{2\pi}{\lambda_P}$, $\lambda_P$ is the wavelength of the probe beam, $\epsilon_0$ is the permittivity of free space, $P(v_x)$ is the probability for a Ca atom to have velocity $v_x$ along the probe laser beam propagation direction, and $\Im(\rho_{01})$ is the imaginary part of the $^{1}S_0 \rightarrow\ ^{1}P_1$ transition density matrix coherence $\rho_{01}$. The Rabi frequency of the probe beam is defined to be $\Omega_P=\frac{\mu_{eg}E_P}{\hbar}$ with $E_P$ being the peak electric-field of the probe beam, $\hbar$ the reduced Planck constant, and $\mu_{eg}$ is the electric dipole moment of the transition. Equation~\ref{eq:alpha1} assumes that the probe laser beam waist is much smaller than the ablation plume extent and that the fractional probe laser beam absorption is small. The change in transmission, $T$, along the direction of propagation of the probe beam, $x$, is $\frac{\partial T}{\partial x}=-\alpha(x,y,z)T$, where $y$ and $z$ are the transverse and longitudinal probe laser beam offsets from the plume center and plume longitudinal origin, respectively.
We assume that, at a given moment, the ablation plume is Gaussian in shape perpendicular to the ablated material's propagation direction, $z$, with radius $r(z)$ at each longitudinal point, so that the atomic beam density as a function of position in the vacuum chamber is
\begin{align}
n(x,y,z) &=n_0\ \textrm{exp}\left[-\frac{x^2 + y^2}{r^2(z)}\right]
\end{align}
where in the experimental layout in Fig. \ref{fig:ExperimentalLayout}, the probe beam propagates along the +$x$-direction. The transmission of the probe beam through the ablation plume is given as the ratio of the probe beam power, $P$, measured by a photo detector to the input laser beam power, $P_0$
\begin{align}
\label{DensityEq7}
T=\frac{P}{P_0}&=\exp{\left[-\int_{-\infty}^{+\infty}\alpha(x,y,z)dx\right]}
\end{align}
\begin{align}
&=\textrm{exp}\left[-n_0(z)\frac{2\sqrt{\pi}k_P r(z) \mu_{eg}^2}{\epsilon_0\hbar\Omega_p}e^{-\frac{y^2}{r^2}}\int_{-\infty}^{+\infty} P(v_x)\mathfrak{I}(\rho_{01})dv_x\right],
\end{align}
where $v_x$ is the Ca velocity along the probe laser beam direction. The peak atom density (units of $m^{-3}$) at a given longitudinal distance from the target, $z$, is then
\begin{align}
n_0(z)=\frac{\epsilon_0\hbar\Omega_p e^{+\frac{y^2}{r^2}}}{2\sqrt{\pi}k_P r(z)\mu_{eg}^2\int_{-\infty}^{+\infty}P(v_x)\mathfrak{I}(\rho_{01})dv_x}\ln{\left(\frac{1}{T}\right)}
\end{align}
and the square of the electric dipole matrix element, $\mu_{eg}$, is
\begin{align}
    \mu_{eg}^2 &= \frac{3 \pi \epsilon_0 \hbar c^3} {\omega_{SP}^3}A_P(2J_P+1)
    \begin{pmatrix}
    J_i & 1 & J_k \\
    m_i & q & -m_k
    \end{pmatrix} ^2
\end{align}
where $A_P\approx2.18\times10^8 ~s^{-1}$ is the inverse lifetime of the $^1$P$_1$ excited state and $J_P=1$ is the total angular momentum of the $^{1}P_1$ level. The $J_i,J_k$ are the total angular momentum, $m_i,m_k$ are the corresponding magnetic quantum numbers for the $i,k$ levels, and $q$ is the spherical basis vector component of the probe laser beam polarization. The following simplification can be made for the $^1 \text{S}_0$-$^1$P$_1$ transition:
\begin{align}
    (2J_P+1)
\begin{pmatrix}
0 & 1 & 1 \\
0 & q & -m_k
\end{pmatrix}^2=1,\ \forall\ q \in \{-1,0,1\}.
\end{align}
We also define $k_P=\frac{\omega_{SP}}{c}$ where $\omega_{SP}$ is the angular frequency of the $S\rightarrow P$ transition and $c$ is the speed of light, reducing the peak density expression to 
\begin{align}
n_0(z)=\frac{1}{6\sqrt{\pi^3}}\left(\frac{k_P^2}{r(z)}\right)\left(\frac{\Omega_p}{A_P}\right)\frac{e^{+\frac{y^2}{r^2}}\ln\frac{1}{T}}{\left[\int_{-\infty}^{+\infty} P(v_x)\mathfrak{I}(\rho_{01})dv_x\right]}.
\end{align}
We insert the analytic expression for $\Im{(\rho_{01})}$ and integrate over all velocities parallel to the probe laser beam, $v_x$, to obtain the Doppler-averaged quantity
\begin{align}
\begin{split}
    \int P(v_x)\mathfrak{I}(\rho_{01})dv_x &= \int_{-\infty}^{+\infty}\sqrt{\frac{m}{2\pi k_b T_{\perp}}}e^{-\frac{mv_x^2}{2k_b T_{\perp}}}
    \\&\left[\frac{\Omega_p\Gamma}{4(\Delta-k_Pv_x)^2+\Gamma^2+2\Omega_p^2}\right]dv_x
\end{split}
\end{align}

\begin{figure*}
    \includegraphics[width=\textwidth]{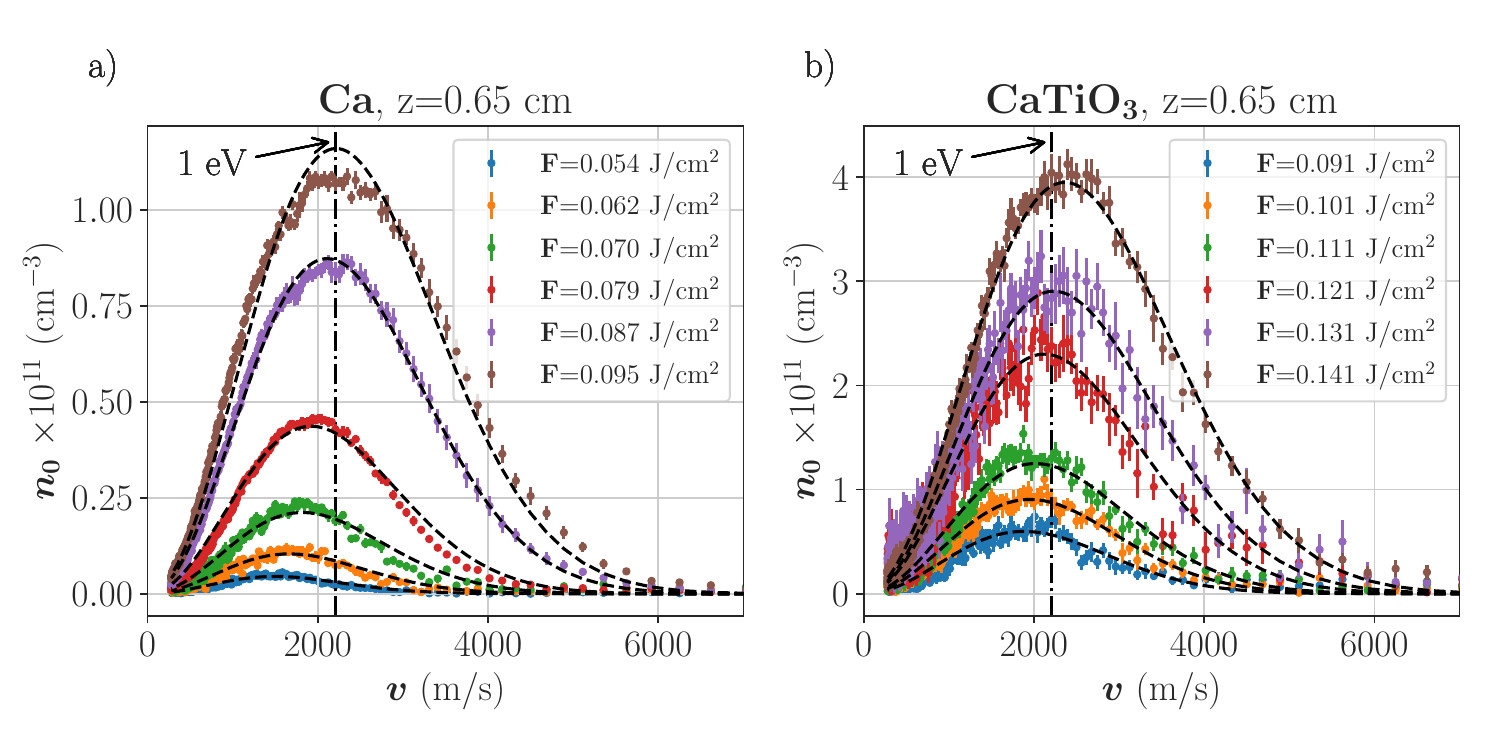}
    \caption{Ablation plume velocity distributions for (a) Ca  and  (b) \titanate\ targets were measured by observing 423 nm probe laser beam transmission (absorption) at varying $z$-distances from the plume center and increasing laser fluence. The velocities are then fit to a Boltzmann distribution to extract plume temperature. For reference, we also plot a vertical line at a Ca velocity of 2196~m/s, which corresponds to $\approx1$~eV of longitudinal kinetic energy.}
    \label{fig:VelocityPlot}
\end{figure*}
\begin{align}
    &=\Omega_p \Gamma\sqrt{\frac{m}{2\pi k_bT_{\perp}}}\int_{-\infty}^\infty\frac{e^{-\frac{mv_x^2}{2k_bT_{\perp}}}}{4(\Delta-k_Pv_x)^2+\Gamma^2+2\Omega_p^2}dv_x,
\end{align}
where $k_b$ is the Boltzmann constant, $m$ is the atomic mass, $\Gamma$ is the total homogeneous broadening of the atomic transition (including excited state decay, collisions, etc.), and $T_{\perp}$ is the plume temperature transverse to the longitudinal ($z$) direction.
Evaluating the Doppler integral for a resonant probe ($\Delta=0$) gives
\begin{align}
\label{Eq:SimplDopplerInt}
\begin{split}
   \int P(v_x)\mathfrak{I}(\rho_{01})dv_x &=\frac{\Omega_p\Gamma}{2k_P}\sqrt{\frac{\pi m}{2k_bT_{\perp}}}\sqrt{\frac{1}{\Gamma^2+2\Omega_p^2}}\exp{\left(\frac{m}{8k_bT_{\perp}}\frac{\Gamma^2+2\Omega_p^2}{k_P^2}\right)}
    \\&\left[1-\textrm{erf}\left(\frac{1}{2\sqrt{2}}\sqrt{\frac{m}{k_bT_{\perp}}}\sqrt{\frac{\Gamma^2+2\Omega_p^2}{k_P^2}}\right)\right].
\end{split}
\end{align}
Equation~\ref{Eq:SimplDopplerInt} can be simplified with the following characteristic velocities we define as

\begin{align}
v_c &\equiv\sqrt{\frac{(\Gamma/2)^2+\Omega_p^2}{k_P^2}}
\end{align}
\begin{align}
    v_{th}&\equiv\sqrt{\frac{2k_bT_{\perp}}{m}}
\end{align}
where $v_c\approx 16$ \ms for $\Omega_p=\Gamma=A_P$ and $v_{th}\approx 350$ \ms for $T_{\perp}$=3000 K.

Applying the result of Eq.~\ref{Eq:SimplDopplerInt}, we obtain a final expression for peak density as a function of probe beam fractional transmission ($T$) at a given probe laser beam vertical offset from plume center ($y$) and longitudinal distance from the target ($z$):

\begin{align}
\label{Eq:peakDensity}
\begin{split}
n_0(y,z)&=\frac{4}{3\pi^2}\frac{k_P^2}{r(z)}\left(\frac{k_P^2 v_c v_{th}}{A_P\Gamma}\right)\exp{\left(+\frac{y^2}{r^2(z)}\right)}\\
&\times\ln\left(\frac{1}{T}\right)\frac{\exp\left(-\frac{v_c^2}{v_{th}^2}\right)}{\left[1-\textrm{erf}\left(\frac{v_c}{v_{th}}\right)\right]}.
\end{split}
\end{align}
In the limits of strong Doppler broadening ($v_{th} >> v_c$), lifetime-limited homogeneous broadening ($\Gamma\approx A_P>>\Omega_P$), probing at the vertical center of the ablation plume ($y=0$), and weak probe absorption ($T \equiv 1 - \delta T$, $\delta T << 1$), Eq.~\ref{Eq:peakDensity} may be simplified to read 
\begin{align}
\label{Eq:peakDensitysimple}
\begin{split}
n_0(y=0, z)\approx\frac{2}{3\pi^2}k_P^3\left(\frac{ v_{th}}{r(z)A_P}\right)\delta T.
\end{split}
\end{align}

\subsection{Ablation Plume Temperature and Density Analysis}

The Ca ablation plume spatial profile is mapped for each target by first translating the probe beam along the $y$-axis at the $z$-positions nearest to the target surfaces while ablating the target at a 1 Hz repetition rate. The vertical center of the ablation plume is found by fitting the peak absorption in $y$ at relatively low ablation laser fluence ($<0.1$ \fluenceUnitsJ). To minimize the ablation laser fluence, the probe beam is moved to the furthest $z$-distance allowed by the beam translation optics (1.2 cm), and the absorption signal is minimized by decreasing the ablation laser power. A full 2-D $y$-$z$ scan is performed for each target at this minimum fluence, providing a 2-D image of the plume absorption for each target. Plume images for each target at similar ablation laser fluence can be seen in Fig.~\ref{PlumeImages}, where we have converted probe beam transmission at each probe laser position into a local, time-averaged atomic number density, $\bar{n}(y,z)$.

We characterize the longitudinal plume velocity for each ablation target by measuring the time-resolved probe beam absorption signal at three different $z$-distances from the target surface at the ablation plume center ($y=0$). We calculate the atomic number density by inserting the time-integrated absorption signal into Eq.~\ref{Eq:peakDensity} and plot representative peak atomic number densities vs. velocity for several pulse energies in Fig.~\ref{fig:VelocityPlot}.

\begin{figure*}[ht]
    \centering
    \includegraphics[width=\textwidth]{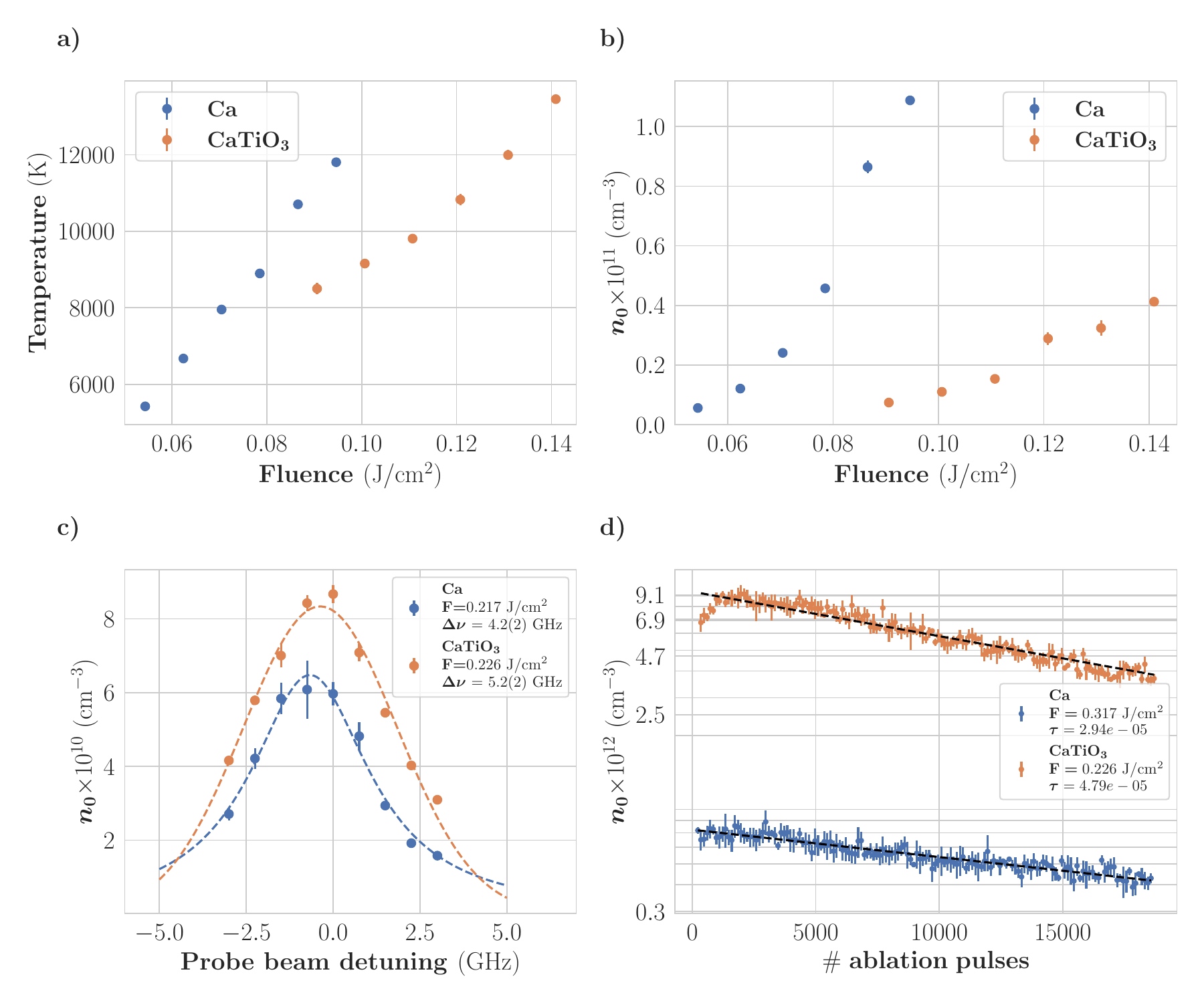}
    \caption{Comparison of a) fit longitudinal atom plume temperature and b) peak plume density vs. laser fluence. c) Peak atom density vs. 423 nm probe beam detuning for both ablation targets, with measured transverse temperatures of 2696(306) K and 4211(334) K for the Ca, CaTiO3 targets respectively. d) Ablation target spot lifetime was measured by increasing the ablation laser fluence to near-saturation of the absorption signal and pulsing at a rep rate of 15 Hz for 20 minutes. The 423 nm absorption was measured in time to measure decrease in atom density for a given spot with increased number of ablation pulses.}
    \label{fig:FrequencyScan}
\end{figure*}

To extract the longitudinal plume temperature, we fit the plume velocity distributions for a range of ablation laser fluences to a Boltzmann distribution, which can be seen in Fig. \ref{fig:VelocityPlot}. We observe most probable velocities for the range of fluences shown in Fig.~\ref{fig:VelocityPlot} of 1502-2215~\ms and 1880-2365~\ms for the Ca and \titanate\ targets respectively, which agrees well with previous pulsed laser ablation experiments \cite{vrijsen_efficient_2019, white_isotope-selective_2022, sheridan_all-optical_2011}. The velocity profiles follow Boltzmann distributions with the exception of  the highest-velocity atoms in the high-pulse-fluence regime. This non-thermal atomic beam component may be the result of collisional effects at higher ablation laser fluence, which have been shown to cause accelerated expansion of the metal vapor from the target surface~\cite{gusarov_target-vapour_2001}. In Fig.~\ref{fig:FrequencyScan}(a), we compare the extracted longitudinal temperature vs. ablation laser fluence for each target, which exhibits a qualitatively linear dependence. Similarly, we plot the calculated peak atomic number density vs. laser fluence for each target in Fig.~\ref{fig:FrequencyScan}(b).

To characterize the transverse plume temperature for both targets, we measure the Doppler-broadened linewidth of the 423 nm probe beam by positioning the beam at $y=0$ near the surface of the targets where the absorption signal is maximized. We then scan the probe laser detuning over a range of $\pm$ 3 GHz from the resonance frequency of the transition, $f_0$, as seen in Fig. \ref{fig:FrequencyScan}(c).
Higher ablation pulse fluences were used in this frequency scan to overcome ablation spot depletion. We extract the transverse temperature by fitting the Doppler-broadened linewidth $\Delta f_{\textrm{fwhm}}$ to a Gaussian, extracting the full width at half maximum, and computing the Doppler temperature $T_{\perp}$ as
\begin{align}
\begin{split}
T_{\perp}=\frac{mc^2}{8k_B\mathrm{ln}(2)}\left(\frac{\Delta f_{\mathrm{fwhm}}}{f_0}\right)^2.
\end{split}
\end{align}
 We measure Doppler-broadened linewidths of 4.2(2) and 5.2(2) GHz, yielding transverse plume temperatures of 2696(306) K and 4211(334) K for the Ca and \titanate\ targets, respectively. 
\begin{figure*}
    \centering
    \includegraphics[width=\textwidth]{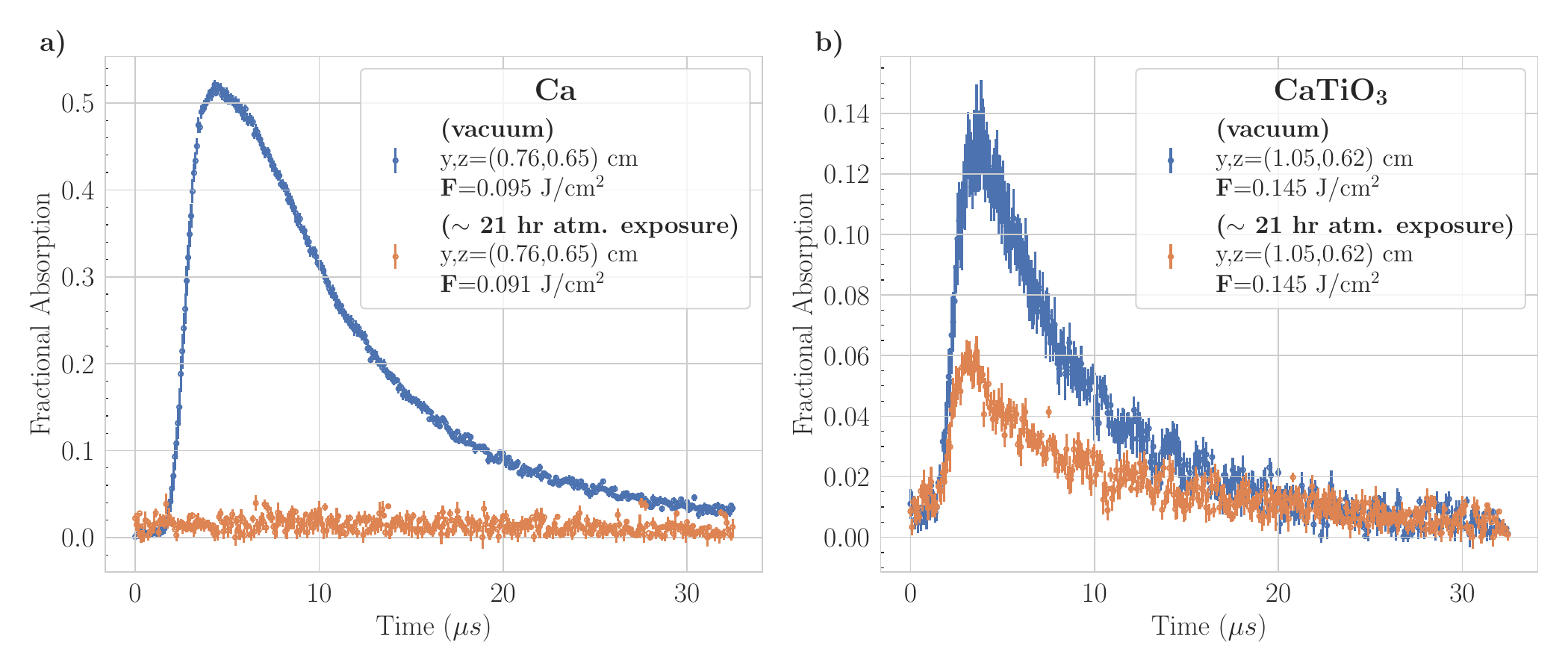}
    \caption{Measurements of fractional probe laser absorption versus time since the ablation laser trigger. Time-dependent probe absorption due to the Ca plumes from each ablation target are shown before (blue points) and after (orange points) 21 hours of atmospheric exposure.}
    \label{fig:AirReactivity_DensityComparison}
\end{figure*}
\subsection{Spot Lifetime}

One concern with the use of ablation targets for loading of trapped ions is the ablation spot lifetime. During the early stages of our experiment, we observed depletion of ablation spots at high fluences while initially setting up the absorption spectroscopy. To quantify the spot lifetime as a function of laser fluence, we place the probe laser beam at the peak plume density vertical position ($y=0$) near the target surface and increase the ablation laser beam fluence to nearly extinguish the probe laser beam transmission signal. We also increase the repetition rate of the ablation laser to the maximum 15 Hz. We collect measurements of the absorption signal over a timescale of approximately 20 minutes. Converting the absorption signal to peak atomic number density, we quantify the peak atomic density as a function of ablation pulse number. As seen in Fig.~\ref{fig:FrequencyScan}(d), we plot the density vs. number of ablation pulses for each target and fit these to a single-exponential function to extract decay constants for the spot lifetimes. We measure similar 1/$e$ decay pulse numbers of $3.4(1)\times10^4$ and $3.2(1)\times10^4$ for the Ca and \titanate\ targets, respectively. By contrast, Osada et al. recently reported \titanate\ exhibiting a longer ablation target lifetime when compared to a pure Ca target for a 1064 nm ablation laser wavelength~\cite{osada_feasibility_2022}.  This could be the result of various differences in our experimental parameters including ablation laser wavelength, angle of incidence, or fluence~\cite{garban-labaune_effect_1982,torrisi_pulsed_2004,hafeez_spectroscopic_2008,torrisi_comparison_2010,zhao_influence_2016}.

\subsection{Target Oxidation}

An additional concern that arises in trapped ion loading sources is oxidation of the source when making vacuum system changes. Atmospheric exposure typically causes oxidation of elemental alkaline earth sources. To test the resistance of \titanate\ to oxidation, we directly compare probe laser beam absorption for Ca and \titanate~targets under similar laser ablation fluences both before and after exposure to atmosphere. We first measure the time-resolved probe beam absorption signal for each target under ideal conditions in-vacuum as discussed earlier in Sec.~\ref{sec:Experiment}. After obtaining the nominal probe beam absorption, we then break vacuum and allow the chamber to sit at atmosphere for approximately 21 hours before pumping back down to similar vacuum pressures and repeating the measurement. Figure~\ref{fig:AirReactivity_DensityComparison} shows the result of these measurements for both targets. In Fig.~\ref{fig:ExperimentalLayout}(a), it is evident that a layer of white oxidation has built up on the Ca target surface after atmospheric exposure, while no visible change was observed for the \titanate\ target. We observe a $\sim$~50\% reduction in probe beam absorption with the \titanate\ target after the atmospheric exposure. This reduction in probe absorption is consistent with observed differences in the yield of different ablation spots on the \titanate\ target. By contrast, we see no evidence of atomic absorption when ablating the Ca target after atmospheric exposure. We also do not observe any noticeable increase in ablated neutral Ca density after $>$500 ablation pulses. These results confirm that \titanate\ is a good candidate for $^{40}$Ca$^+$ experiments requiring frequent vacuum changes, or in complex experimental apparatuses where the vacuum system cannot be immediately pumped after oven assembly (e.g. some cryogenic systems). 
\begin{figure}
    \centering
    \includegraphics[width=\columnwidth]{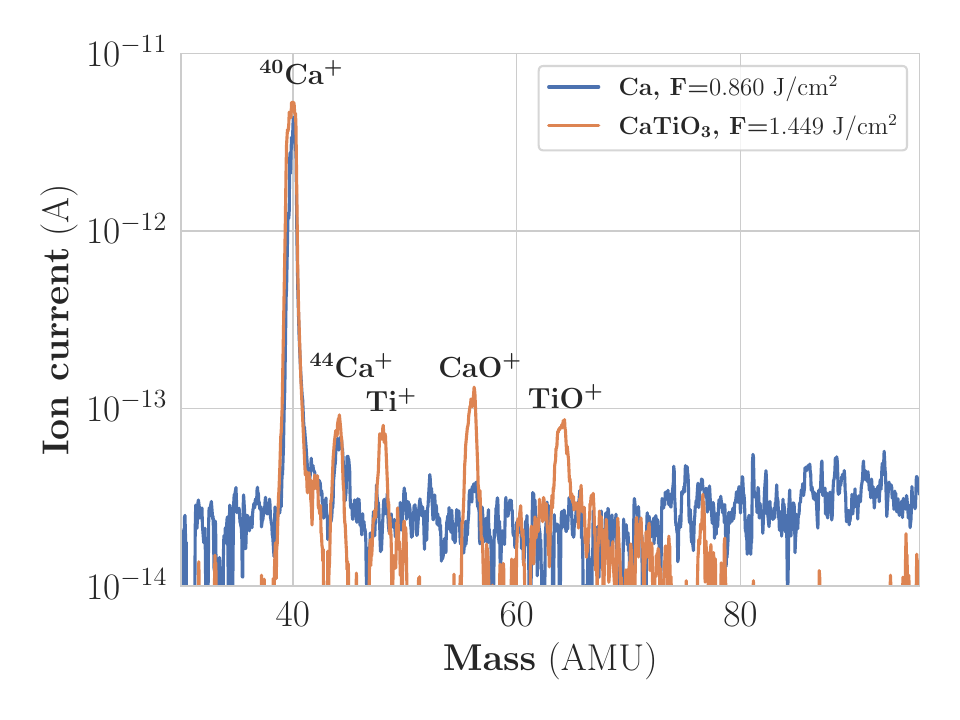}
    \caption{Comparison of mass spectra for each ablation target at an ablation pulse repetition rate of 1 Hz.}
    \label{fig:IonCurrent_Comparison}
\end{figure}

\subsection{Ion Production}
Finally, we measure the direct ion generation for each target based on mass spectrometry using a commercial residual gas analyzer (RGA). The RGA is attached to one of the vacuum chamber radial ports so that its inlet is directly facing the ablation targets at a distance of $\sim10$~cm. For these measurements, the RGA ionizer filament and external repeller are disabled to minimize background ion currents. A mass spectrum (see Fig.~\ref{fig:IonCurrent_Comparison}) is taken with the RGA while pulsing the ablation laser at a repetition rate of 15 Hz. We convert the pressure reading of the RGA into ion current based on the specified instrument conversion factor to record the relative number of different ion species produced during the ablation process. For reference, we measure a $^{44}$Ca/$^{40}$Ca ratio of 0.02(1) for both targets, which is in good agreement with the natural abundance value of 0.02152(5) \cite{meija_isotopic_2016}. For the \titanate\ target, we additionally observe Ti$^+$, CaO$^+$, and TiO$^+$ within the given scan upper limit of 100 atomic mass units. We note that our ablation vacuum system has no electrodes for guiding the ablated ion flux to the RGA, therefore the absolute ion currents reported in Fig.~\ref{fig:IonCurrent_Comparison} do not necessarily represent the total ion flux from the targets.

\section{Conclusions}

We have characterized the phase space distribution of $^{40}$Ca neutral atomic beams produced by pulsed laser ablation of Ca and \titanate\ targets through absorption spectroscopy. Measuring time-dependent absorption of a 423~nm probe laser beam, we observe peak atomic densities from each ablation target on the order of $10^{11}-10^{12}$~cm$^{-3}$ at ablation laser fluences in the range of 0.1-0.3 \fluenceUnitsJ. We quantify the ablation plume temperature and velocity for a range of ablation laser pulse fluences, with longitudinal plume temperatures of 5423-11804 K for Ca and 8499-13453 K for \titanate. The measured longitudinal plume velocities for the same range of fluences are 1502-2215~\ms and 1880-2365 for the Ca and \titanate\ targets, respectively. Transverse plume temperature extracted from measurements of the Doppler broadening of the Ca absorption are below the measured longitudinal temperatures, reflecting the directionality of the atom plumes.  

Our results highlight the benefit of choosing a perovskite ablation target for ion loading by directly comparing the effects of atmospheric exposure on atomic density yield for each target. We measured far less reduction in the neutral atomic flux from the \titanate\ target after 21 hours of atmospheric exposure, while we observed no evidence of atomic flux from the Ca target after visible surface oxidation of the target. The continuous 21-hour exposure is
representative of the \titanate target longevity when subjected to multiple atmospheric exposures during vacuum changes which can be as short as 1-2 hours. In addition, we quantified the spot lifetime for each target at reasonably high ablation laser fluences (0.18-0.30 \fluenceUnitsJ) where we saw 20-30\% decrease in neutral atomic flux after tens of thousands of ablation pulses. We anticipate that the ablated atom plume densities and temperatures reported here, when combined with ionization models, may be used for quantitative estimates of ion trap loading rates in more restricted trap geometries. Future work could include measurements of ablated Ca density within an ion trap apparatus and the relative photoionization loading efficiency of Ca$^+$ obtained from different target materials.

\backmatter


\bmhead{Acknowledgements}
The authors acknowledge Jonathan Jeffrey for contributions to the ion detection experimental setup. We also thank Creston D. Herold for useful discussions and review of this manuscript. The authors acknowledge funding from  Office of Naval Research Grant N00014-20-1-2427 and the Georgia Tech Research Institute.

\section*{Declarations}
Data will be made available upon reasonable request.

\bibliography{references}

\end{document}